4# Enabling Disaster Resilient 4G Mobile Communication Networks

Karina Gomez, Leonardo Goratti, Tinku Rasheed and Laurent Reynaud*Abstract*—The 4G Long Term Evolution (LTE) is the cellular technology expected to outperform the previous generations and to some extent revolutionize the experience of the users by taking advantage of the most advanced radio access techniques (i.e. OFDMA, SC-FDMA, MIMO). However, the strong dependencies between user equipments (UEs), base stations (eNBs) and the Evolved Packet Core (EPC) limit the flexibility, manageability and resiliency in such networks. In case the communication links between UEs-eNB or eNB-EPC are disrupted, UEs are in fact unable to communicate. In this article, we reshape the 4G mobile network to move towards more virtual and distributed architectures for improving disaster resilience, drastically reducing the dependency between UEs, eNBs and EPC. The contribution of this work is twofold. We firstly present the Flexible Management Entity (FME), a distributed entity which leverages on virtualized EPC functionalities in 4G cellular systems. Second, we introduce a simple and novel device-to-device (D2D) communication scheme allowing the UEs in physical proximity to communicate directly without resorting to the coordination with an eNB.

*Index Terms*—Direct Mode, Distributed Cellular Entities, Public Safety Networks, Disaster resilient communications## I. Introduction

4G-LTE networks are designed in the way that the base stations (eNBs) depend on many other local and regional technical entities located within an Evolved Packet Core (EPC) to ensure their proper operation. This strong dependency between the access and the core networks limit the flexibility, manageability and resiliency of 4G-LTE systems. In fact, physical destruction of network components has been identified as the most common cause of telecommunications failures in recent disasters. For example, during the recent Japan tsunami in 2011, a total of approximately 1.9 million fixed communication lines and 29,000 base stations were damaged [1]. Additionally, disrupted communication links between the access and core network affect the communications even if the network elements are perfectly working from the hardware and software point of view. Thus, problems caused by physical destruction are also more severe and are likely to last longer than the problems caused by network congestion for instance. Another limitation for resilient communications in 4G-LTE networks is the strong dependency between UEs and the access networks. In fact, UEs not only need to communicate in traditional cellular fashion, but they also need to communicate directly in case the network infrastructure is temporarily unavailable or if the operating conditions prevent reliable communication links. In the new Release-12 of 3GPP (Third Generation Partnership Project), device-to-device (D2D) communications is being specified and will help to overcome the strong dependency between UEs and the radio access network (RAN) [2]. Essentially, we consider D2D communications associated to relief problems arising from dead spots where the eNB signal might be not available or however very weak for prolonged durations and to enable communications in peer-to-peer mode.

The EPC is defined by 3GPP with the goal of providing simplified all-IP core network architecture to efficiently give access to various services. Using the user and control plane mechanisms, EPC supports a set of specialized functions such as to enforce access control, perform user authentication and implement a number of application services, just to name a few. Consequently, if even a single entity inside the EPC fails the whole network operations are affected. In order to reduce events of failure as much as possible, EPC entities employ complex techniques to provide the highest levels of reliability that are needed for mobile operators to serve hundreds or thousands of users. Typically, multi-blade server setups with two-way redundant components and complex and accurate system monitoring are deployed. Due to these factors, the installations or replacement of hardware entities requires careful planning over long timescales and the intervention of trained staff. Consequently, installing and operating such entities may cause significantly higher Capital Expenditures (CAPEX) and Operating Expenses (OPEX), which also reduce the resiliency of the network during crisis situations.

We focus on designing a software architecture and a set of distributed protocols able to provide higher flexibility, manageability and resilience of 4G communication networks while ensuring the levels of reliability similar to operational cellular network infrastructures in crisis and disaster situations. In order to achieve our goal of augmenting 4G networks with critical infrastructure capabilities, we propose a novel component of the LTE system architecture called flexible management entity (FME), which is based on the idea of EPC entities virtualization that entails the deployment of customized services and resource management solutions

"The research leading to these results has received partial funding from the EC Seventh Framework Programme (FP7-2011-8) under the Grant Agreement FP7-ICT-318632".

Karina Gomez, Leonardo Goratti and Tinku Rasheed are with CREATE-NET, via alla Cascata 56D, 38123 Trento, Italy (e-mail: name.surname@create-net.org).

Laurent Reynaud is with Orange Labs, 2 Avenue Pierre Marzin, 22307 Lannion, France (e-mail: laurent.reynaud@orange.com).

locally at eNBs, thus reducing or eliminating in some cases, the dependence on physical EPC entities. We argue that devising means to embed the most fundamental EPC operations at the RAN side using virtual LTE-EPC entities is a fundamental step towards obtaining high-performing mobile network architectures, as discussed in [3] [4]. Secondly, we also discuss a complementary solution to enable resilient mobile network communications, based on an innovative scheme for D2D networks where a UE broadcasts beacon frames to allow other UEs to recover connectivity at least in proximity (i.e. short range distance). The main advantages of the proposed D2D scheme compared to other solutions present in literature [5][6][7][8] are that (i) the necessary operations to discover, establish and maintain D2D communications are totally independent of eNB and core network entities, and (ii) we propose using LTE uplink channels (PUCCH and PRACH) for running our protocol, making it easy to incorporate in LTE specifications. Thus, the flexible and resilient 4G-LTE architecture model is based on an embedded eNB-EPC model, which can be fully standalone and operational with the support of virtual end-to-end physical core infrastructures, and UEs able to operate independently from the eNB using D2D communication modes whenever this is necessary. In this way we reshape the traditional cellular network operations in order to move towards virtual-distributed architectures reducing drastically the dependency between UEs, eNBs and EPC enabling critical disaster resilience.

## II. STATE OF THE ART

### A. EPC Virtualization

As mentioned in the introduction, decoupling the tight dependency between EPC functionalities and the RAN is an evolutionary step allowing higher resilience. In particular, this holds true in cases of crisis scenarios and post-disasters management. Following this idea, the work done in [3] starts a discussion on alternative ways of network ownership. The authors introduce the concept of virtual operators arguing that this allows the creation of more flexible cellular network environments while ensuring more efficient architectures. Similar concepts are also investigated in [9], where a reconfigurable mobile network architecture is proposed for flexibility and reconfigurability. 3GPP has also recognized the importance of supporting network sharing by means of virtualization through the different Releases 6-12 where a set of technical specifications and architectural requirements are defined [10].

The design and implementation of network virtualization substrate (NVS) for effective virtualization of wireless resources in cellular networks is introduced in [11]. The authors analyze the 3GPP ongoing efforts in RAN sharing enhancements and introduce a concrete implementation of a RAN sharing base station virtualization solution for the LTE system. In [12], authors introduce the Distributed Mobility Management Entity (DMME) that implements mobility management for the next-generation of cellular systems using a distributed LTE Mobility Management Entity (MME). Effectively, DMME is a scalable and cost-effective drop-in replacement of the LTE-MME. In [13] authors argue that carrier networks can benefit from advances in computer science and pertinent technology trends. The paper introduces a virtual network management framework for implementing network architectures based on a software-defined networking (SDN) approach.

In [3],[9],[10] the focus is on sharing the core network, whereas of the work in [11] mainly focuses on wireless resource sharing. On the contrary, the key innovation we propose consists of moving key EPC functions more closer to the RAN to enable a distributed control and management of 4G networks. Our work extends the idea of virtualization and decentralization to all EPC functions in order to achieve a more flexible, manageable and resilient 4G-LTE network architecture as required in disaster and crisis management.

### B. D2D Communication Protocols

D2D communications have been recently discussed in 3GPP in the context of so called Proximity Services (ProSe) study item [2]. It is worth mentioning that LTE-based device-to-device communications have recently received significant attention by both academia and industry. In [5] an intra-cluster D2D retransmission scheme with optimized resource utilization is presented. The scheme can adaptively select the number of cooperative relays performing multicast retransmissions and give an iterative sub-cluster partition algorithm to enhance retransmission throughput, and the D2D cluster formation procedure is under the supervision of the eNB. Similar approaches are studied also in [6], in which authors propose a D2D discovery and link setup procedure. In the proposed procedure the $UE_{TX}$ broadcasts a beacon for performing network discovery. After receiving a beacon, $UE_{RX}$ requests to its eNB to set up a D2D link with $UE_{TX}$. Thus, eNB participates in establishing the D2D link allocating the resources for setting up the temporal link between $UE_{TX}$ and $UE_{RX}$.

A different approach is presented in [7] in which a D2D server coordinates the establishment of D2D communication links by maintaining and tracking the capabilities of the D2D UEs as well as interacting with the MME for performing D2D bearer's setup procedures. Similarly, in [8] the authors introduce the network protocol and architecture for LTE-A based D2D communications, where the UE play the role of D2D enabler while the packet data network gateway (P-GW) is the D2D coordinator. To establish and manage the D2D bearer, the eNB, P-GW and MME exchange messages for creating the D2D bearers, thus providing D2D-based data offloading.

To the best of our knowledge, the majority of the D2D protocols available from the literature prescribes establishing the D2D link between UEs with the coordination/supervision of eNB [5][6] or core network entities [7][8]. We noticed that this approach affects the resiliency of the D2D network in all cases in which either the eNB or some core network entity is unavailable, thus implying that the D2D network cannot function properly. We address this issue by proposing a D2D communication scheme that works independently from the access and core network entities.

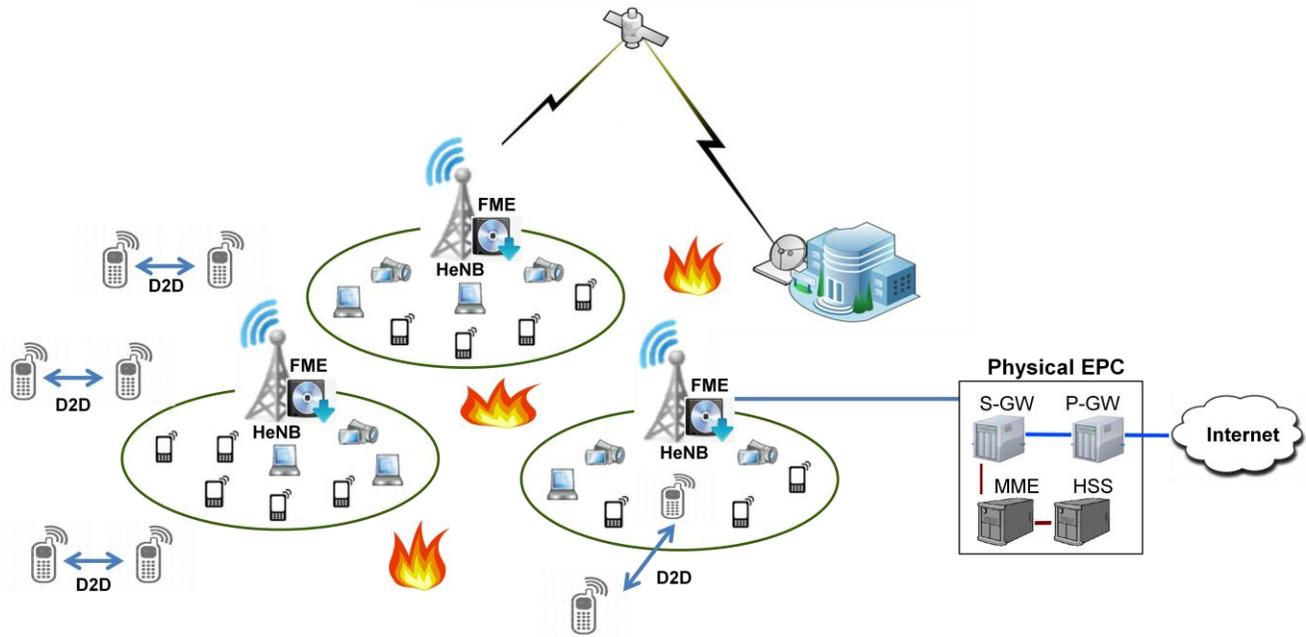

Fig. 1. FME and D2D application scenarios in resilient communications.

## III. SCENARIOS AND MOTIVATIONS

Crisis scenarios are often characterized by damaged network elements or severely impaired communication links between network nodes and entities. In these cases, robust architectures are required in order to keep the network services running with low or possibly even no impact to the end users. However, the current 4G-LTE network architecture is strongly influenced by the need of monitoring the user traffic constantly. This increases the dependencies between hierarchical network entities and greatly reduces the resilience and robustness of the whole system in case of crisis events. In 3GPP Release 11 specific restoration procedures were developed for the EPC entities. The study includes the MME, Serving Gateway (S-GW), and P-GW recovery mechanisms after a failure with and without restarting such entities [14]. However, these procedures are only detecting failures at a software level. Thus the current 4G network is not designed for supporting hardware entities failures, which is the case of the disruption caused by the occurrence of disasters that may seriously affect the provision of services to the end users.

In order to reduce the dependency between network entities and to make the network more robust, we propose a flexible 4G-LTE architecture embedding the most fundamental EPC functions inside the eNB, which we term as Hybrid-eNB (HeNB). Fig. 1 illustrates an example of a flexible 4G-LTE network deployment, in which the HeNB is designed to work totally isolated (if needed) from the physical EPC. Furthermore, the HeNB supports particular functionalities of the physical EPC to enable an autonomous behavior for the provision of connectivity and services to the users or at least internal communication for the UEs. The HeNB can include wired or wireless connectivity (IEEE 802.11, 802.16, Satellite, etc.) for ensuring a link to the physical EPC. Thus, the proposed flexible 4G-LTE architecture also implies dynamic radio environments, where only a subset of HeNBs manages the communications with the physical EPC when it is required.

To design resilient 4G networks, we move from a highly centralized system to a distributed concept as depicted in Fig. 2. As we can observe, the combination of physical and virtual network entities coexist enabling distributed operations while the D2D communication mode allows in establishing an ad hoc LTE network. To allow the virtual and physical EPCs to coexist in dynamic scenarios, specific routing and topology management mechanisms are required. The benefits of virtualizing EPC and enabling D2D communication mode are mainly:

- *Dependency Reduction and Flexible Network Deployments:* Enabling both the UEs working independently from the eNB as well as the eNB to operate in a standalone way greatly reduces the interdependency between hierarchical network elements of 4G networks. This also allows exploring new network architectures like for example, mobile eNB supported by wireless backhauling.

- *Resilient Communications:* Making the access and core entities able to work independently from each other will allow the network to have a high level of resiliency. Thus, in the worst scenarios in which physical entities are out of service in terms of hardware or software, the network is able to s*elf-organize* and *self-operate*.

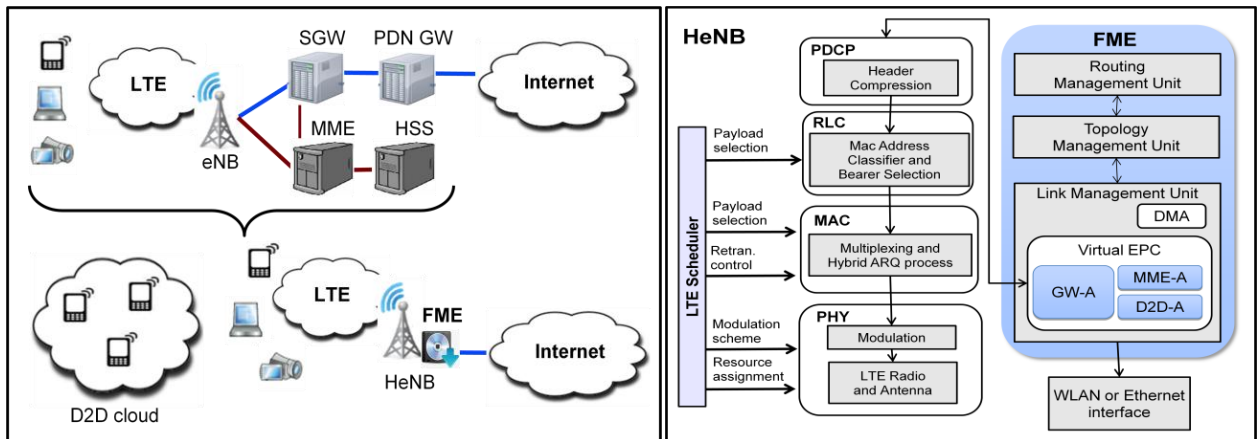

Fig. 2. Resilient 4G-LTE Architecture, software elements and interfaces.

- *Flexible Network Operations and Resource Management:* Since the virtual EPC functions is less dependent on proprietary hardware, adding functions or updating/modifying protocols is faster and easier to be performed.

IV. FLEXIBLE MANAGEMENT ENTITY DESIGN

In this section, we introduce FME, where the main objectives are i) to embed the most fundamental EPC operations at the eNB side and ii) make the coexistence possible between virtual and physical EPCs. The FME software elements and interfaces are shown in Fig. 2, it is composed of the following elements:

*A. Virtual EPC*

Virtual EPC supports specific EPC functionalities that give the HeNB the freedom to operate autonomously by providing connectivity and other services to the users. Specifically, the virtual EPC functionalities are supported with three autonomic functions, namely Gateway-Agent, Mobility Management Entity-Agent and D2D agent.

*1) Gateway-Agent (GW-A)*

It manages all the mechanisms implemented for supporting the basic functionalities of EPC from the user plane point of view. This agent is responsible for guaranteeing the proper operation of the HeNB when it is disrupted or disconnected from the physical EPC or when the physical EPC does not exist. The GW-A essentially consists of a dynamic code repository, which is used as a protected code execution environment for the virtual EPC entities. This technique guarantees that if a specific server located in the physical EPC is temporarily unavailable (e.g. S-GW or P-GW), GW-A runs a function able to act as a surrogate server. Consequently, the GW-A is responsible for:

- Deciding which functions have to be activated to guarantee the integrity of operation of the HeNB. Prioritizing the functionalities that run over the GW-A according to the scenario and the available resources in the HeNB.

- Providing connectivity of the UEs to external packet data networks and reestablishing and performing the handover of all the user plane functions to the physical EPC when it is required.

It is worth noting that the complexity of GW-A is dependent on the number of EPC functionalities that it executes.

*2) Mobility Management Entity-Agent (MME-A) and D2D-Agent (MME-A)*

The mechanisms implemented for enabling control plane functionalities are supported by the MME-A. It manages and stores UEs information regarding their identities, mobility state and security parameters (see Fig. 2). In fact, the MME-A interacts with LMU (see next subsection) for supporting the creation of virtual X2 (vX2) interfaces for interconnecting HeNBs. In general, the MME-A is responsible for:

- Interacting with external MME-A and RMU for performing the handover procedures between virtual EPCs and synchronizing periodically with the physical EPC the UE user plane contexts.

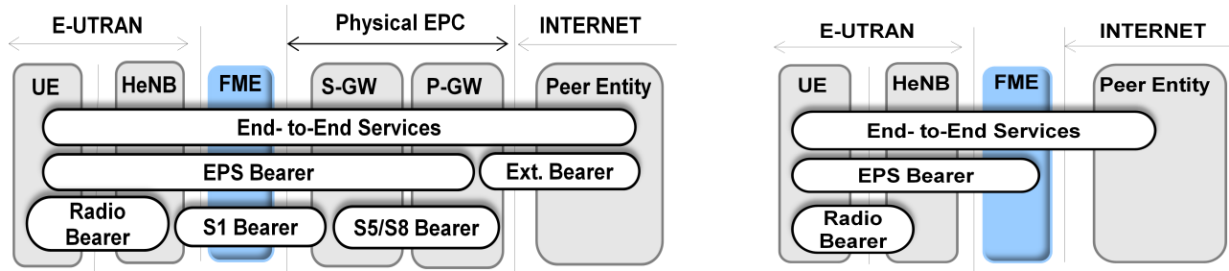

Fig. 3. FME role with and without physical EPC.

- Reestablishing and performing the handover of all the control plane functions to the physical EPC (MME specifically).

It is worth pointing out that the D2D-A is responsible for managing all the mechanisms implemented for D2D mode communications as well as storing the D2D context information. Furthermore, notice that the D2D-A is meant to provide the necessary functions for allowing D2D mode communications amongst the UEs inside the HeNB coverage area.

*B. Link Management Unit (LMU)*

This unit manages the MAC and PHY communication of the new wireless interfaces supported by the HeNB in order to communicate with the physical EPC. This unit is also responsible for encapsulation/de-encapsulation of all the messages exchanged between HeNBs and the physical EPC regardless of the available technology supported by HeNBs (i.e. satellite or WiFi). This procedure mainly creates a tunnel between the HeNB and the physical EPC. This unit is responsible for maintaining the direct or multi-hop link between the HeNBs and the physical EPC in order to 1) create a virtual-S1 (vS1) interface as well as to tunnel S1 into vS1 interface, and 2) create vX2 interfaces for interconnecting HeNBs when the MME-A performs UEs handover. Due to the highly challenging hypothesis on HeNB mobility patterns and unpredictable changes of the wireless channel conditions (between HeNBs and physical EPC), the probability of frequent link disruption between HeNBs and the physical EPC or external networks can be quite high in certain scenarios. Consequently, the FME supports adapted and autonomic mechanisms in order to avoid information loss, thus reducing the probability of service interruptions. For this purpose, the disruption management agent (DMA) is used.

*C. Routing and Topology Management Units*

RMU is responsible for routing packets in the network and maintaining active routes between each HeNB and the physical EPC. The Topology Management Unit (TMU) is instead responsible for topology optimization of the HeNBs. These units allow a dynamic topology where the HeNB can always have a link with physical EPC if it is required. A high level overview of the role of FME in both scenarios with and without physical EPC is illustrated in Fig. 3 while Fig. 4 shows an overview of the FME handshake messages and summarizes the FME units and agents interactions.

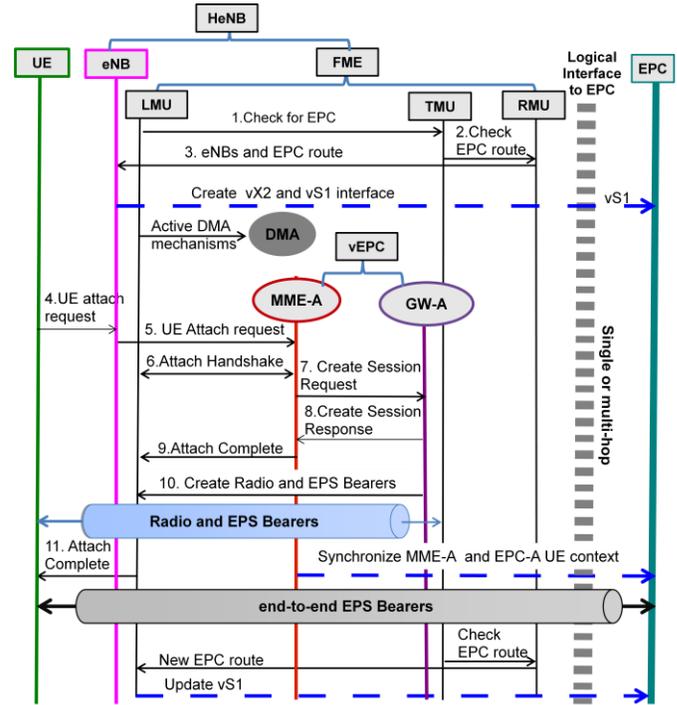

Fig. 4. FME handshake messages overview.

Specifically, the messages exchanged involve the following phases:

**Creating the vX2 and vS1:** When the HeNB is activated, it first sends an interlayer discovery message to the TMU asking for information about the HeNBs and physical EPC present in the network (messages 1, the requested information includes HeNBs and physical EPC Identifiers and routes). Then, the TMU sends a request message to the RMU in order to activate the routing protocols and obtain the best routes for the HeNBs and physical EPC (messages 2). Thus the interlayer discovery and request messages are answered by the TMU and RMU in order to update the LMU with the required information (messages 3). Finally, using this information LMU maintains a table with HeNBs and physical EPC ID and the routes and create the virtual vX2 and vS1 interfaces. Additionally, HeNB also enable the DMA mechanism in order to support the network disconnection.

**Creating the Radio and Evolved Packet System (EPS) Bearer:** When a UE requests the association with the HeNB (messages 4), it interacts with MME-A and GW-A for completing the request and to be able to serve the UE. Specifically, HeNB sends an attach request message to the MME-A (messages 5). Then, the MME-A and HeNB exchange attach handshake messages (messages 6) while the MME-A and GW-A exchange session request messages for completing the UE attached procedures with the HeNB (messages 7-8-9). When, the attached and session creation procedures are completed, the EPS bearer is created by the GW-A and the UE is notified about it (messages 10-11). At this point, the UE can be served and the intra-cell data or voice communications can be performed.

TABLE I
D2D CONSIDERATIONS

| D2D OPEN QUESTIONS | Possibilities |
|---|---|
| What conditions to trigger D2D? | • Network congestion<br>• Network failure<br>• Out coverage situation<br>• Intermittent Links |
| What resources for D2D? | • Time-Division Duplex (TDD)<br>• Frequency-Division Duplex (FDD)<br>• Uplink (SC-FDMA)<br>• Downlink (OFDMA) |
| Who is deciding D2D? | • Base Station (eNB)<br>• User equipment (UE)<br>• EPC entities<br>• Negotiation between UE and eNB |
| Which frequencies are available for D2D? | • License<br>• Unlicensed |

TABLE II
D2D PROTOCOL OVERVIEW

| LTE UPLINK CHANNELS | Role in D2D protocol |
|---|---|
| Physical Uplink Control Channel (PUCCH) | • D-Beacons from b-UE<br>• Authentication and Association responses |
| Physical Random Access Channel (PRACH) | • Authentication requests to the b-UE |
| Reserved PUSCH slots | • Association requests to the b-UE<br>• D2D links reservation Request<br>• D2D links reservation Response |
| Physical Uplink Shared Channel (PUSCH) | • D2D data exchange |

**Creating end-to-end EPS Bearer:** Once the EPS bearer is created, the MME-A and GW-A send the UE context information to the physical EPC in order to synchronize the UE information within the whole network. Then an end-to-end EPS Bearer can be activated with the interaction of virtual and real EPC. At this point, the inter-cell data or voice communications can be performed.

Thus, the FME units (LMU, TMU and RMU) interact in order to create and maintain the vX2 and vS1 interface while the FME agents (GW-A and MME-A) create and maintain the bearers of the different services that are required for serving the UEs (refer Fig. 4). Consequently, associated UEs can be immediately served for inter-cell/intra-cell communication. To conclude, the FME is a software solution that distributes several core functionalities at the access side allowing the eNB to keep the network services running with low or no impact to the end user in case of network disruption.

## V. D2D COMMUNICATION SCHEME

While FME reduces the dependency between access and core networks, mechanisms for allowing the UEs to operate independently from the access network are also required. In this way, 4G (and future 5G communication infrastructures) networks will leverage resilient communications similar to a tactical communication network. Currently, D2D communications modes are being investigated by 3GPP, in which public safety communications are supported. The most relevant questions to be answered when designing a D2D communication scheme are summarized in Table I. We propose a D2D communication protocol answering those questions in the way that the UE itself will be the enabler, coordinator and manager of the D2D network even without any preliminary interaction with the access network. Thus, when connectivity with the eNB is lost or non-existing during a *time-of-interruption (ToI)*, a selected UE is responsible for establishing and managing the D2D network using single-carrier frequency division multiple access (SC-FDMA) as modulation format for transmission and reception. The adoption of a ToI is necessary to avoid undesired ping-pong effects. The main advantage of using SC-FDMA is clearly to protect cellular downlink communications from interferences and to preserve energy. We also make the assumption that frequency division duplexing (FDD) is used. Notice that investigating the impact of ToI in setting up D2D networks is out of the scope of this paper, although relevant in general. The salient features of the proposed D2D communication protocol are summarized below.

### A. D2D Network Discovery

In case the eNB is not available for at least a ToI, a UE starts the procedure for creating or joining a D2D network. Thus, a UE, which is denoted as *b-UE,* evaluates the possibility of broadcasting *direct beacon frames (D-beacons).* In principle, any UE could transmit a beacon frame, but before doing that UEs must listen to the channel for at least two *D-beacon interval (TD)* in order to i) check whether there is already a D2D network around and ii) avoid collisions and interference. We refer to *D-beacon interval (TD)* as the period between two consecutive beacon transmissions. The D-beacon interval is supposed to be any multiple integer of the LTE radio frame duration (i.e. 10 ms). In case a beacon is not received, the UE is allowed to start broadcasting its own beacon frame. The transmission of D-beacons by the b-UE has the purpose of replacing the signals from the eNB for other UEs. The first advantage of this solution is to enable a timely discovery of the neighborhood. Some of the information periodically conveyed by the D-beacon frame is for example the D2D network identification and the identity of all the UEs connected to that particular D2D network. In order to exploit the available uplink resources and structure, the physical uplink control channel (PUCCH) is exclusively used by the b-UE to transmit D-beacons (refer Fig. 5) and for example to reply to network association requests of other UEs. For a single LTE subframe, the PUCCH is located in one RB at or near the edge of the system bandwidth and in a second RB at the opposite edge to exploit frequency diversity, together called as the PUCCH region. Note that the number of PUCCH regions depends on the bandwidth of the LTE system.

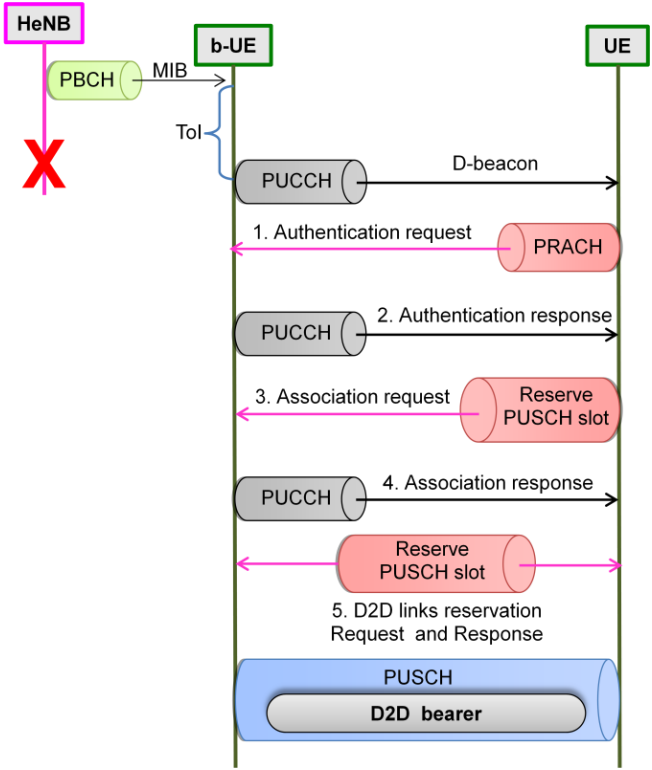

Fig. 5. Handshake messages for creating D2D Networks.

TABLE III
SIMULATION PARAMETERS FOR CHANNEL

| Link | LTE UL/DL | D2D | WiFi UL/DL |
|---|---|---|---|
| Channel Model | Clarke's Fading 20 fading paths | Free Space | Free Space |
| Modulation | 16-QAM 1/3(SISO) | QPSK | BPSK/QPSK/QAM |
| α | 2.2 | 2.1 | 2 |
| Transmission Power | UE: 23 dBm eNB: 30 dBm | 23 dBm | 23 dBm |
| Central Frequencies | 700 MHz | 700 MHz | 5 GHz |
| Antenna Gain | 3/3 dB | 0/0 dB | 3/3 dB |
| Receiver Sensitivity | UE: -107.5 dBm eNB: -123.4 dBm | -107.5 dBm | -85 dBm |
| Thermal Noise | UE: -104.5 dBm eNB: -118.4 dBm | -174 dBm/Hz | -110 dBm |

Notice that in case the eNB is available, the authorization or initialization of D2D mode communications can also be controlled by the D2D-A, which is available in the FME architecture shown in Fig. 2. Consequently, for the purposes of coverage area extension or offload traffic from HeNB, the FME allows UEs inside the coverage area to setup D2D mode communications.

### B. D2D Network Establishment

After receiving the *D-beacons* broadcasted by the *b-UE*, the UEs can join the D2D network. Thus, the UEs trying to establish a connection with the b-UE shall follow the four-way handshake defined in standard LTE specifications and that define random access channel operations (RACH), which is shown in Fig. 6. Logical RACH functions are carried over the physical RACH (PRACH) slots that are all over the structure of the D-beacon interval defined by the b-UE. We assume that PUCCH is also used by the b-UE to respond to association/authentication requests of the UEs. To join the D2D network, as in the normal LTE procedure, UEs randomly select one out of 64 Zadoff-Chu (ZC) preamble signatures thus mitigating the risk of potential collisions. In fact, we assume that multiple independent ZC sequences can be correctly decoded by the b-UE. We propose that the response of the b-UE shall follow over the PUCCH rather than the physical downlink shared channel (PDSCH) as in standard LTE but still in contention-less mode. Furthermore, we also allocate reserved slots in the physical uplink shared channel (PUSCH) to carry the third message sent by the UEs during the four-way handshake, as well as to reserve resources for the purpose of exchanging data or setting up voice calls (see Fig. 5). It is worth pointing out that reservation of resources for data and voice does not involve the b-UE but rather takes place among peer entities. The preamble contention resolution shall follow standard LTE procedure, being the b-UE assigning D2D network identifications. At this stage we also envisage room to apply backoff schemes (e.g. Binary exponential backoff) in order to reduce collisions over the same preamble sequence. Notice that the role of the b-UE can optionally be rotated among UEs for the purpose of preserving energy. The framing structure summarizing our proposal is shown in Tab. II

### C. D2D Network Disassociation

In order to enable timely discovery of the eNB, we suppose that periodically all UEs (including the beaconing terminal) shall make an attempt to detect the synchronization signals and the master information block (MIB) sent by the eNB over the physical broadcast channel (PBCH). In fact, we assume that if a particular UE manages to recover connectivity with the eNB, resources allocated to that D2D network must be relinquished. In this way FME objectives are complemented by enabling the UEs with the capability of forming ad hoc topologies, thus reducing drastically the dependency between user equipment, access and core networks elements.

## VI. PERFORMANCE ANALYSIS

We present here a brief evaluation of the advantages arising from the use of FME and D2D for flexible 4G networks. The implemented HeNB node model is depicted in Fig. 2 while the simulation scenario is shown in Fig. 1. The network scenario was simulated in OMNeT++ using LTE simulator [15]. The simulation model is adapted to support the implementation of FME as previously explained. We consider a scenario over a surface of 2 km$^2$ and we assume that a natural disaster cut off communication links between the HeNB and EPC. In the disaster area 3 HeNBs are working with a distance of approximately 1 km between each HeNB. We assume that a physical EPC is located approximately 1.5 km away from the disaster area. Furthermore only a single HeNB has direct connectivity with the physical EPC. The HeNB backhaul network is created using an ad hoc WiFi link. Notice that WiFi was selected for the purpose of the simulations and it is necessary stressing the fact that we are not proposing WiFi as a solution for the backhaul. In fact, one of the main objectives here is to validate the role of FME in managing dynamic scenarios. For the sake of D2D, we assume that M=75 UEs are outside the coverage area of the HeNBs, thus they are allowed to setup D2D mode communications. Table III summarizes the parameters used for modeling LTE, WiFi and D2D links parameters. The HeNB transmission power is set to 30 dB for achieving a Microcell with a coverage radius of 500 m. The cell is configured with a TDD duplex scheme with 10 MHz bandwidth using 50 full resource allocations and 16-QAM modulation. Inside the cell, 250 UEs are uniformly distributed with transmit power of 23 dBm. The UEs move inside the served area following a random waypoint mobility model with a uniformly distributed speed between (0.2-0.7) m/s. The following multimedia applications are simulated:

- 40 intracell calls and 5 UL/DL real-time video streaming between UEs and the Video Server for each cell,
- 20 intercell calls between cell-1/cell-2 and cell-2/cell-3 (calls between UEs of different cells are termed as cell-x/cell-y).

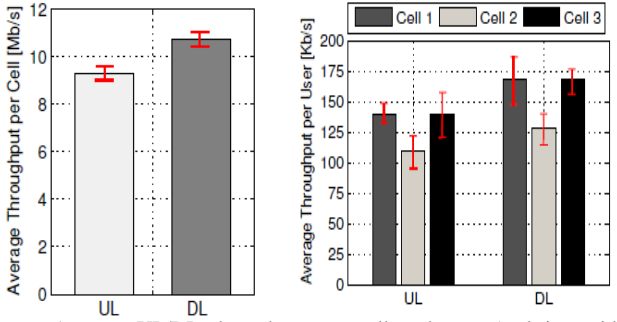

Fig. 6. Average UL/DL throughput per cell and user (real-time video streaming and intracell calls) for 4G-LTE-based network scenario. The cell is configured with (D|S|U|U|D|D|S|U|U|D) TDD-LTE mode using 16-QAM in 10 MHz with full resource allocations.

The intercell calls were simulated using enhanced voice services (EVS) codec encoded at 64 kb/s and real-time video streaming was simulated using H.264 Codec encoded at 384 kb/s. The applications run in parallel for 600 seconds. Results presented in this section are averaged over 10 simulation rounds plotted with the 95% confidence interval. While for D2D, a path-loss exponent equal to 2.1 is used (similar to free-space propagation). We assume a D-beacon interval TD of 80ms and a D-beacon duration equal to 10% overhead of the whole TD. Fig. 6 shows the average UL/DL throughput per cell and user. We chose the slot configuration number 1 where the slots dedicated for DL/UL are the same (sharing frequency) and the same cell configuration for all the cells. As it can be seen, the achieved throughput in UL and DL are similar (around 10 Mb/s) as both use similar configuration distinguished only by the amount of overhead introduced. While the average UL/DL throughput achieved for each served user is similar to the cell-1 and cell-3, the average UL/DL throughput for cell-2 is less than for cell-1/cell-3, since cell-2 has more users to serve. We conclude saying that the virtual version of the EPC is able to perform functionalities of routing for serving the cells and allowing intracell and intercell calls in the HeNB networks. Furthermore, it allows video streaming from external networks passing through the physical EPC.

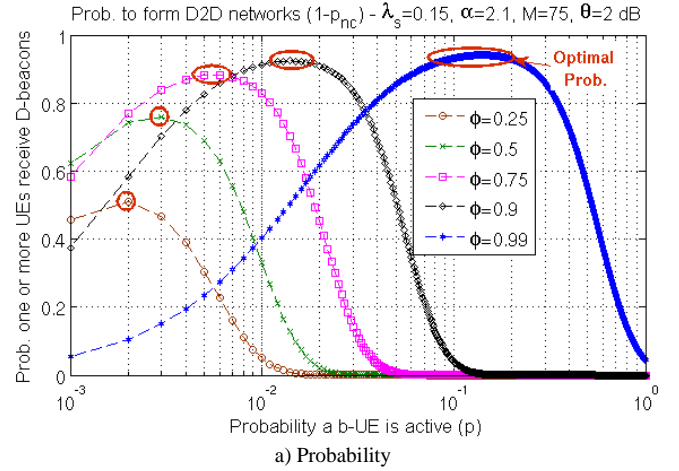

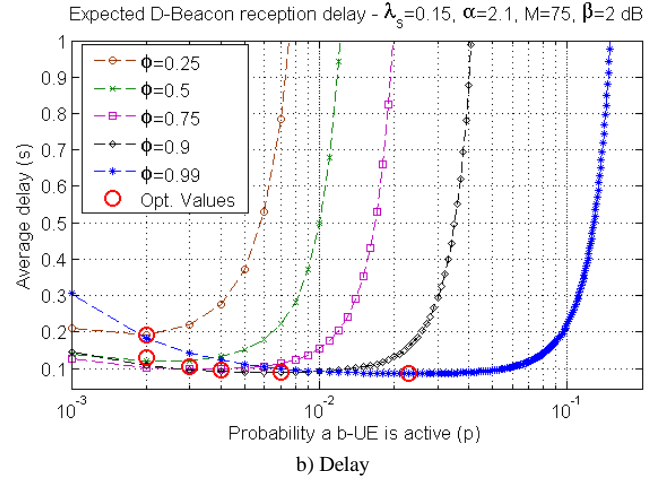

Fig. 7. a) Probability at least a D-beacon is received correctly by UEs and b) the D-beacon reception Delay while varying the probability a b-UE is active, assuming a total of M=75 UEs.

In the case of D2D, Fig. 7 (a) shows the probability to form D2D networks in non-coverage region. The results are obtained assuming that UEs are scattered over the two-dimensional Euclidean space according a homogeneous Poisson point process of intensity $\lambda_s$, as deeply explained in [16]. In the results, the parameter $\Phi$ denotes the fraction of regular UEs with respect to the total (i.e. non-beaconing UEs). Essentially the analysis investigates the connectivity aspect of the D2D network and the delay that arise from the D2D network formation. As visible in the figures, this allows identifying optimal values of the probability a b-UE is actively sending D-beacon frames. The optimum value of probability tends to shift from left to right as the density of UEs is increased (vice-versa the density of b-UEs is decreased) and more than a single value the curves show almost a narrow region. Decreasing the number of active b-UEs implies that we need more of them transmitting beacons in order to ascertain connectivity in the region. Fig. 7 (b) shows that the D-beacon reception delay is in agreement with the behavior observed in the connectivity analysis. Consequently, the proposed D2D protocol shows the capability of improving the communication resilience of UEs outside the network coverage.

## VII. CONCLUSIONS

We described FME, a novel architectural solution to realize simplified and virtualized EPC functions, which is an enabler for deploying and managing resilient eNBs that are capable of standalone and autonomous operations. This reduces the capital and operational costs, as well as time and effort in re-deploying a multi-service, multi-band inter-operable and integrated network infrastructure for specific applications, including emergency communications. FME is an example of a software-defined solution for running the virtual EPC inside the ends for the provisioning of services even when the physical EPC fails or is unavailable. Aiming at achieving more standalone network operations, we also proposed a novel D2D communication protocol endowing the LTE-enabled mobile user equipments with ad hoc capabilities. The advantage of the proposed protocol is the flexibility of starting D2D mode communications without any interaction with the radio access network.

## BIOGRAPHIES

**Karina Mabell Gomez Chavez** (karina.gomez@create-net.org) was born in Chillanes, Ecuador. She received the engineering degree (cum laude) in Electronic and Telecommunication Engineering from the National Polytechnic School in Ecuador, in 2006. She received her Master degree in Wireless Systems and Related Technologies from the Turin Polytechnic, Italy, during 2007. In year 2007, she joined FIAT Research Center, becoming part of the Infomobility-Communication and location Technologies. In July 2008, she joined iNSPIRE Area at Create-Net, working on several projects She received her PhD degree in Telecommunications from the University of Trento, Italy, during 2013. During her PhD she conducted various industry internships and research visit at Orange Labs in Lannion-France, Telekom Innovation Laboratories, Berlin-Germany, School of Electrical and Computer Engineering, RMIT University, Melbourne-Australia. Since beginning of 2013, she is part of the FuN Area at Create-Net. She has several patents on protocols for next generation of wireless networks and has published her research in important journals and conferences.

**Leonardo Goratti** (leonardo.goratti@create-net.org) received his PhD degree in Wireless Communications in 2011 from the University of Oulu-Finland and his M.Sc. in Telecommunications engineering in 2002 from the University of Firenze-Italy. From 2003 until 2010, he worked at the Centre for Wireless Communications (CWC) Oulu-Finland first as a researcher and then as a PhD student. His research interests cover Medium Access Control (MAC) protocols for wireless personal/body area networks and wireless sensor networks, as well as routing protocols for sensor networks. His research interests cover also UWB transmission technology and 60 GHz communications. From 2010 until early 2013 he worked on MAC protocols for cognitive radios and spectrum sharing techniques at the European funded Joint Research Centre (JRC) of Ispra, Italy. Recently he joined the Research Centre CREATE-NET Trento-Italy where he is currently working in the European project ABSOLUTE on LTE-based device-to-device communications in the context of public safety scenarios.

10**Tinku Rasheed** (tinku.rasheed@create-net.org) is a Senior Research staff member at Create-Net. Since May 2013, he is heading the Future Networks R&D Area [FuN] within Create-Net. Before joining Create-Net in December 2006, Mr. Rasheed was research engineer with Orange Labs R&D from May 2003 until November, 2006. Mr . Rasheed received his Ph.D. degree from the Computer Science Department of the University of Paris-Sud XI., in 2007. He completed his M.S. degree in 2003 from Aston University, U.K. specializing in Telecommunication engineering and his bachelor degree in 2002 in Electronics engineering from University of Kerala, India. Dr. Rasheed has extensive industrial and academic research experience in the areas of mobile wireless communication and data technologies, end-to-end network architectures and services. He has several granted patents on distributed protocols for wireless networking and has published his research in major journals and conferences.

**Laurent Reynaud** (laurent.reynaud@orange.com) is a senior research engineer and expert for the Future Networks research community at Orange. Specialized into the performance of agile infrastructures for challenging environments, his topics of interest include wireless multi-hop, multipath and large scale routing techniques as well as the various issues related to QoS in wireless networks. After receiving his engineering degree from ESIGETEL at Fontainebleau in 1996, he acquired a significant experience regarding the development and deployment of distributed software in the context of telecommunications, through successive positions in the French Home Department in 1997, in Alcatel-Lucent from 1998 to 2000, and in Orange since 2000. He participated to many French, European and international cooperative research projects. He co-authored many conference and journal articles, holds a dozen international patents and regularly serves as TPC member or reviewer for several ACM/IEEE/IFIP conferences and journals.